\documentclass[review]{elsarticle}

\usepackage{lineno,hyperref}
\modulolinenumbers[5]
\journal{Physica A}

\usepackage{graphicx}
\usepackage{dcolumn}
\usepackage{bm}
\usepackage{longtable}
\usepackage{array}
\usepackage{booktabs} 
\usepackage{multirow}
\usepackage{color}
\usepackage{rotating}

\def\R{{\mathbf{R}}} 
\def\Z{{\mathbf{Z}}}
\def\P{{\mathcal{P}}}
\def\G{{\mathcal{G}}}
\def\E{{\mathbf{E}}} 
\def\n{{\mathbf{n}}}
\def\p{{\mathbf{p}}}
\def\d{{\mathbf{d}}}

\let\eps=\epsilon

\def\PA{${\rm PA}(m,A)$}
\def\UniS{${\rm SF}(\gamma,m)$}
\def\me{${\rm ME}(\p)$}
\newtheorem{defn}{Definition}
\newtheorem{algorithm}{Algorithm}

\begin{document}

\begin{frontmatter}

\title{Exactly scale-free scale-free networks}

\author[china,perth]{Linjun Zhang}
\ead{zlj11112222@gmail.com}
\author[perth]{Michael Small\corref{callme}}
\cortext[callme]{Corresponding author}
\ead{michael.small@uwa.edu.au}
\author[perth]{Kevin Judd}
\address[china]{Department of Statistics and Finance\\ University of Science and Technology of China, Hefei, P.R.China, 230026}
\address[perth]{School of Mathematics and Statistics\\The University of Western Australia, Crawley, WA, Australia, 6009}

\begin{abstract}
  Many complex natural and physical systems exhibit patterns of
  interconnection that conform, approximately, to a network structure
  referred to as scale-free. Preferential attachment is one of many
  algorithms that have been introduced to model the growth and structure
  of scale-free networks. With so many different models of scale-free
  networks it is unclear what properties of scale-free networks are
  typical, and what properties are peculiarities of a particular growth or
  construction process. We propose a simple  maximum entropy process which provides
  the best representation of what are typical properties of scale-free
  networks, and provides a standard against which real and algorithmically
  generated networks can be compared.  As an example we consider preferential attachment and find that this particular growth model does not yield typical realizations of scale-free networks.  In particular, the widely discussed ``fragility'' of scale-free networks is actually found to be due to the peculiar ``hub-centric'' structure of preferential attachment networks. We
  provide a method to generate or remove this latent hub-centric bias ---
  thereby demonstrating exactly which features of preferential attachment
  networks are atypical of the broader class of scale-free networks.  We
  are also able to statistically demonstrate whether real networks are
  typical realizations of scale-free networks, or networks with that particular degree distribution; using a new surrogate
  generation method for complex networks, exactly analogous the the widely
  used surrogate tests of nonlinear time series analysis.
\end{abstract}


\end{frontmatter}


\section{Introduction}
The notion of scale-free networks has been around for a
while~\cite{dP76}. The introduction of the
\emph{preferential attachment} (PA) algorithm for generating random
scale-free graphs was a significant step in understanding the properties
of scale-free networks, and the physical processes that create them \cite{aB99}. PA
has spawned a good deal of subsequent algorithms and  analysis.
The purpose of this paper is to highlight the straightforward fact that not all scale-free networks
have the same properties, and that algorithms, like PA  (for example), do not capture the
full richness of scale-free networks, nor do they necessarily display 
properties that may be termed typical of all scale-free networks.

To achieve our aim, we first briefly recall in this introduction the
principal processes that have been proposed to describe and generate
scale-free networks, and indicate deficiencies of these processes when employed as models of
typical scale-free networks. We then propose a \emph{maximum entropy
  process} that provides an unbiased sample of the set of all scale-free
networks. A \emph{maximum entropy process} provides a better
representation of expected properties of scale-free networks, both in
terms of richness and typicality. A \emph{maximum entropy process} also
provides a unbiased standard against which other processes that generate
scale-free networks can be compared. While a great deal of this maximum entropy process reduces to simple edge-switching, it is the application of this process to achieve maximum entropy realisations that is important.

In Sec.~\ref{sec:results} we make a careful comparison of PA with our
unbiased standard to illustrate how PA has a
significant bias in the structural properties of the networks it
generates. We demonstrate it has a \emph{hub-centric} bias.
In Sec.~\ref{sec:applications} we use a surrogate data approach to examine a
selection of real-world networks claimed to be scale-free, and analyze in
what sense these networks are typical of scale-free networks and how they
differ.

\subsection{Scale-free networks from processes}

In this section we briefly review the notion of scale-free networks, and
the principal models of the physical processes that generate scale-free
networks. These models can be broadly divided into \emph{growth} models
and \emph{configuration} models.

Scale-free networks are usually identified by the histogram of node
degrees having a power-law tail. Many naturally occurring networks{\footnote{For a sample of the data we use here, see {\tt http://vlado.fmf.uni-lj.si/pub/networks/data/} .}} have
been identified as having a scale-free property: citation and
collaboration networks~\cite{dP65,aW04}, airline
networks~\cite{rG04}, protein-protein interaction~\cite{hJ01}, metabolic
pathways~\cite{jD05}, the world-wide web and internet~\cite{aH13}.

To understand the formation of scale-free networks various models have
been proposed to mimic the physical or conceptual processes that build and
shape these networks. One of the first, proposed by Barab\'asi and Albert,
is \emph{preferential attachment}~\cite{aB99}, which is a
restatement of the process described by de Solla Price~\cite{dP76} as
a model of observed scale-free networks of citations~\cite{dP65}.

Preferential attachment is an \emph{unchanging}, \emph{additive} growth
process, where nodes of a fixed degree are added to the network, with
links preferentially attached to existing nodes depending on their degree;
usually proportional to the degree. Although this was suggested as a model
of the process that grows the world-wide web (seen as hyperlinks between
webpages), it was recognized that there is an additional \emph{aging}
process (AOL is replaced by Facebook, yahoo looses popularity to Google)
so that attachment preferences \emph{change} as the network growth
{proceeds}~\cite{rX04,pK01,dP02}.

Other processes can be introduced into a network growth model, such as
\emph{shuffling} parts of the network, and \emph{deletion} of links and
nodes~\cite{rX04, kJ13}, such that these actions do not change
the underlying scale-free property of the networks produced. Moreover, 
scale-free networks can be produced by processes that
are not growth processes. The most commonly considered of these are the so-called {\em configuration models}.  Configuration models proceed by choosing nodes
to have prescribed degrees, then connecting these together to form a
scale-free network~\cite{mM95}; although care is needed to ensure
the networks obtained satisfy other expected properties, such as, being
simple (no self-links or multiple edges between nodes) and
connected~\cite{tB06, pC12, kJ13}. In addition to the problems with multplie edges and self-links \cite{kJ13}, configuration models  do not provide a well-founded sampling --- in the sense of achieving maximum entropy realisations \cite{gB09}.

\subsection{Not all scale-free networks are the same}

With so many different processes generating scale-free networks, the
question arises: do they generate the same type of networks? The simple answer is
no. Many differences in the scale-networks generated by different
processes have been noted; some particular differences are outlined in the
following.

Consider a preferential attachment process \PA{} that {{attaches}} nodes of
degree~$m$ with the probability of linking a new node to a node of
degree~$k$ is $k+A-1$, where $A\geq1$ is a constant called the
\emph{initial attractiveness}~\cite{sD00c}. The minimum degree~$m$
clearly effects the \emph{robustness} of the networks generated, and it
effects other properties too~\cite{kJ13}. It can be shown that for \PA{}
the power-law tail has an exponent $\gamma=2+A/m$. This type of
preferential attachment process restricts the power-law exponent to
$\gamma>2$. Many real-world networks exhibit $\gamma<2$. In such cases the graphs have been termed {\em dense}~\cite{cG11} and generating such graphs has been seen as problematic. This led Del Genio and colleagues to conclude eponymously  to the contrary.  However, we recently provided an algorithm that easily find dense graph --- the apparent paradox with the claim of \cite{cG11} is resolved when one observes that \cite{kJ13} searches over the space of all networks while \cite{cG11} starts with the constraint of viable graphs of a fixed degree sequence. Hence, both \cite{fF13} and \cite{kJ13} proposed that these network growth models can be modified to shape networks and allow such
power-laws. 

When other network statistics are considered (such as diameter, node-degree
assortativity, clustering coefficients), then the differences between
scale-free networks found in the real-world and generated by different
models becomes more apparent.

For example, consider the node-degree assortativity of scale-free networks
generated by preferential attachment. These networks have low
assortativity, but many natural networks are found to have very high
assortativity~\cite{mS08}. Although these natural networks are not
small-world (although they are scale-free typically estimates of the exponent are much smaller than observed for most processes that generate scale-free networks:  $\gamma<2$), it can be shown that other processes can generate scale-free
networks with high node-degree
assortativity~\cite{skinny2}. Newman~\cite{mN03} observed systematic
biases in the assortativity of real-world scale-free networks:
technological networks tended to be dis-assortative, social networks
assortative. He also showed that networks generated by preferential
attachment have a definite bias compared to real-world networks; on the
other hand, \emph{altruistic attachment} does not~\cite{pL11}. Other issues
with the assortativity of preferential attachment networks ~\cite{rX04,xu3} and scale-free networks more generally \cite{sJ10} have been
observed previously.

It has been noted that growth models have systematic
biases (of network statistics) relative to configuration
models~\cite{dC01,sD00c}.  However, while configuration models can successfully alleviate that bias, it is not clear that they provide a sampling of the appropriate distribution. In particular, networks of growth models
usually only attain the scale-free property asymptotically, and small
networks display systematic biases~\cite{xu3,mS08};  configuration
models remove this particular growth-based bias, but it is not clear at what expense. For this reason we seek a maximum entropy realisation from the ensemble of viable graphs \cite{gB08}.

\subsection{Maximum entropy processes}

In this section we aim to define a process that generates scale-free
networks in as pure form as possible, that is, without any of these biases. This
process is a \emph{maximum entropy} process, which can serve as a standard
to which other processes can be compared to both recognize and understand
the nature of their particular biases.

As previously noted, \emph{scale-free} is usually taken to mean that the
node-degree histogram has a power-law tail. We need to make this
definition more precise. Let $\G$ be the set of connected simple graphs
with $N$ nodes, and let $\n(G)=(n_1,\dots,n_{N-1})$, $n_k\in\Z$, be the
degree histogram of~$G\in\G$, that is, $n_k$ is the number of nodes of
degree~$k$.

\begin{defn}\label{defn:tail}
  If $\p=(p_1,\dots,p_{N-1})$, where $p_k\in[0,1]$ and
  $\sum_{k=1}^{N-1}p_k=1$, then $\p$ has a \emph{power-law tail} if
  $p_k\propto(k-\alpha)^{-\gamma}$ for $k\geq\beta$, where
  $\alpha,\beta\in\Z$, $\gamma\in\R$, $\alpha,\beta\geq0$ and
  $\gamma>1$.
\end{defn}

In this definition $\alpha$ shifts the power-law
tail relative to~$k$,  $\beta$ is where
the power-law tail is deemed to begin, and 
$\gamma$ is the power-law of the tail.
For the special case of preferential attachment \PA{}, the key parameters
$\alpha$, $\beta$ and $\gamma$ are implicitly defined by $m$ and~$A$.

\begin{defn}\label{defn:sfgraph}
  $G\in\G$ is a \emph{scale-free graph} if $\n(G)\simeq N\p$, where $\p$
  has a power-law tail.
\end{defn}

Although Defn.~\ref{defn:sfgraph} is apparently what most researchers appear to mean by \emph{scale-free}, what \emph{approximately equal} means
in~$\n(G)\simeq N\p$ is not always clear. It is usually taken to mean that
the form $(k-\alpha)^{-\gamma}$ \emph{fits} $\n(G)$, in the sense of
least-squares curve-fitting of $\log{}n_k$ against $\log{}k$, or a
multinomial fit of $\n$ against~$\p$.

Of interest are processes generating independent random scale-free
graphs. A stationary stochastic process~$\P$ generating independent random
graphs in~$\G$ is equivalent to selecting graphs according to a fixed
probability mass function $\P:\G\to[0,1]$, where
$\sum_{G\in\G}\P(G)=1$. Write $G\sim\P$ to denote that $G\in\G$ is
selected according to process~$\P$. (Since the process is completely
defined by its probability mass function, we use the same symbol~$\P$ to
denote the process and its probability mass function.)

\begin{defn}\label{defn:sfprocess}
  A process~$\P$ \emph{generates} scale-free graphs (on average), if
  \begin{equation}
    \label{eq:sf}
    \E[\,\n(G) \mid G\sim\P\,] = N\p,
  \end{equation}
  where $\p$ has a power-law tail.
\end{defn}

Such a process generates \emph{scale-free graphs}, according to
Defn.~\ref{defn:sfgraph}, with the notion of \emph{approximately equal},
$\n(G)\simeq{}N\p$, implicitly defined by~$\P$. For example, one can
compute the variance 
\[\E\left[\,\sum_{k=1}^{N-1}(n_k-Np_k)^2 \Biggm| G\sim\P\,.\right],\]
or compute the probability of a histogram~$\n(G)$ deviating from $N\p$ in
some norm by more than a threshold~$\eps$, that is,
$\Pr(\|\n(G)-N\p\|>\eps \mid G\sim\P)$.

Preferential attachment \PA{}  is one example of such a process; a network is
grown by random preferential attachment until a graph with $N$ nodes is
obtained. This process implicitly defines the probability mass function
for $G\in\G$, which depends on~$m$ and on how the seed network is
chosen. Other growth models (with link ageing, initial attractiveness,
shuffling) will usually obtain different probability mass functions.

A difficulty with growth processes, including \PA{}, is that although they
satisfy Defn.~\ref{defn:sfprocess} of scale-free processes, the
power-law~$\gamma$ is implicitly defined. In some cases (although not for
PA) it is not clear that $\gamma$ is necessarily the same for all
graphs generated for fixed growth parameters. Depending on the initial
seed graph, it is possible to imagine growth processes which would
converge to different power-laws. In contrast, one of the advantages of
configuration models is the mean histogram~$N\p$ is prescribed: the
process configures nodes with pre-defined degrees into a
network~\cite{mM95}.

Even if two processes generate scale-free graphs by
Defn.~\ref{defn:sfprocess}, or have the same power-law~$\gamma$, or have
the same expected histogram~$N\p$, the expected value of other graph
statistics such as assortativity, clustering, diameter, may be quite
different --- as noted above. If we are to understand the nature of scale-free graphs, and the nature of
the processes that generate them, then it is useful to establish some
\emph{benchmark} or \emph{standard} against which processes are
compared. To some extent \PA{} has served that propose, principally
because it is a simple growth process. Unfortunately, it is not a
satisfactory standard for many reasons we already indicated: it has a
limited range of~$\gamma$ ({{\em typically} $\gamma$ is constrained to be between $2$ and $3$}; for $\alpha\gg m$ and $N\rightarrow\infty$ one obtains $\gamma\rightarrow 3$); $\p$ and $\gamma$ are implicitly defined; and, is subject to
well known biases of several basic network measures (such as assortativity).

We propose that the best standard against which to compare processes that
generate scale-free networks is a \emph{maximum entropy}
process~\me{} with prescribed mean histogram~$N\p$. A maximum entropy
process samples all graphs with mean histogram~$N\p$ with the least bias
or emphasis of structures and features that are not common features of all
such graphs.

\begin{defn}\label{defn:maxentropy}
  The \emph{maximum entropy process}~\me{} uniformly samples $\n$
  by the multinomial distribution of~$\p$, and uniformly samples graphs
  with the histogram~$\n$. That is, $\Pr(G \mid G\sim\P_{{\rm ME}(\p)}) =
  \Pr(G\mid\n)\Pr(\n\mid\p)$, where
  \begin{equation}
    \label{eq:uniform}
    \Pr(G\mid\n) = \frac{1}{\left|\left\{G\in\G\mid{}\n(G)=\n\right\}\right|}
  \end{equation}
  and
  \begin{equation}
    \label{eq:multinomial}
    \Pr( \n \mid \p) = N!\prod_{k=1}^{N-1}\frac{p_k^{n_k}}{n_k!}.
  \end{equation}
\end{defn}
The process~\me{} generates connected graphs with maximum entropy from the specified degree sequence
partitions all graphs in~$\G$ into equivalence classes by histogram~$\n$,
and uniformly samples these~(\ref{eq:uniform}), hence, achieving maximum
entropy within these equivalence classes.  Moreover, the sampling of~$\n$ uses a
multinomial distribution~(\ref{eq:multinomial}), which is the maximum
entropy process for selecting a histogram with frequencies~$\p$. Hence, even
though the degrees of nodes within a graph are not independent, the
degrees of nodes of successive randomly generated graphs behave as though
they were independent. That is, maximum entropy is a consequence of selecting degree histogram by multinomial sampling and thence uniform sampling from within that degree histogram. This is equivalent to the work presented by Bianconi \cite{gB08, gB09} for selecting maximum entropy for a specified ensemble. For us, degree distribution and connectivity constraints dictate the ensemble of interest.

For details on how to efficiently implement~${ME(\p)}$, see
Algorithm~\ref{alg:MEp} in \ref{App:mainalg}. A \emph{maximum
  entropy} process~${ME(\p)}$ can be implemented by as a Monte Carlo
Markov Chain (MCMC)~\cite{kJ13}, however, a naive MCMC implementation
could converge slowly and not be effective for large
graphs~\cite{pD12, pD13, yC05}. The key ideas of
Algorithm~\ref{alg:MEp} are that the equivalence class of graphs with the
same degree histogram can be sampled uniformily by a process of \emph{edge
  switching}~\cite{aS11, rT81}, and a cannonical graph with a
given degree histogram can be constructed by the Havel-Hakimi
process~\cite{vH55, sH62}. Combining these key ingredients with a test
for connectivity~\cite{kJ13} provides a sufficient algorithm.

Following Bianconi \cite{gB08,gB09} we define the maximum entropy scale free network ensemble as follows.
\begin{defn}\label{defn:UniS}
  Let $SF(\gamma,m)$ denote the maximum entropy process $ME(\p)$, where
  $\p$ has a power-law tail~$\gamma$, with $\alpha=\beta=m$ and $p_k=0$
  for $k<m$.
\end{defn}
In the following $SF(\gamma,m)$ will serve as our standard against which
scale-free graphs and scale-free graph generators are compared.

\section{A hub of hubs is more than a rich club}
\label{sec:results}

In this section we use algorithm~\ref{alg:MEp} to explore what are the
typical properties of scale-free networks, and moreover, infer that PA
networks exhibit an atypical ``hub-centric'' structure. Power-law
distributions alone imply that the degree distribution has a long tail and
some nodes in the tail must have high degree, but the presence of ``hubs''
implies something more and are often said to ``hold the network
together''. What we observe in the following is something more than
this. There are three properties to which we refer with the term
``hub-centric'' (which we define later in an algorithmic sense): (i) the
distribution of hubs throughout the network, (ii) their interconnection,
and (iii) their connection with low-degree nodes. The presence or absence
of these three properties determine, to a very great extent, many of the
global properties of a scale-free network. Moreover, we observe that PA
networks exhibit properties which are atypical of the broader class.

However, before moving to the conclusion that the difference between these
two types of networks is due to the hub-centric properties of PA networks,
we first present a numerical study of the various main measures of network
geometry: assortativity, clustering coefficient and diameter. We apply
these measures to various different families of scale-free networks to
highlight the wide variation which is possible. We will also explore other
prominent features --- most importantly: (i) the robustness to targeted
worst-case attacks (i.e. ``attack vulnerability''
\cite{aR00}) on hubs; and (ii) a detailed motif
analysis.

For our comparison experiments we generate four types of PA$(m,A)$
networks, with $m=1,2,3,4$ and $A=0$. For given $N$ and~$m$ we generate a
sample from PA$(m,A)$ of~$M$ graphs~$G_i$, $i=1,\dots,M$. From a degree
histogram~$\n(G)$ we use the following formula~\cite{aC09} to
estimate the power-law tail exponent~$\gamma$ of~$G$, as
\begin{equation} \label{eq:est_g} \widehat\gamma(G) = 1+N
    \left(\sum_{k=1}^{N-1} n_k\log\frac{k}{m-\frac12}\right)^{-1}.
 \end{equation}
Letting $\gamma=\frac1M\sum_{i=1}^M\widehat\gamma(G_i)$ we use
Algorithm~\ref{alg:MEp} and generate an sample of~$M$ graphs from
\UniS{}. Results reported in the following are for $N=2000$, $m=1$, $2$, $3$,
$4$, and $M=40$. Computation with other values produced comparable
results. In the following subsections we examine each of the network
properties outlined above, starting with motif rank and then robustness.

\subsection{Comparison of motif rank }

\begin{figure}
\begin{center}
\includegraphics[width=0.95\textwidth]{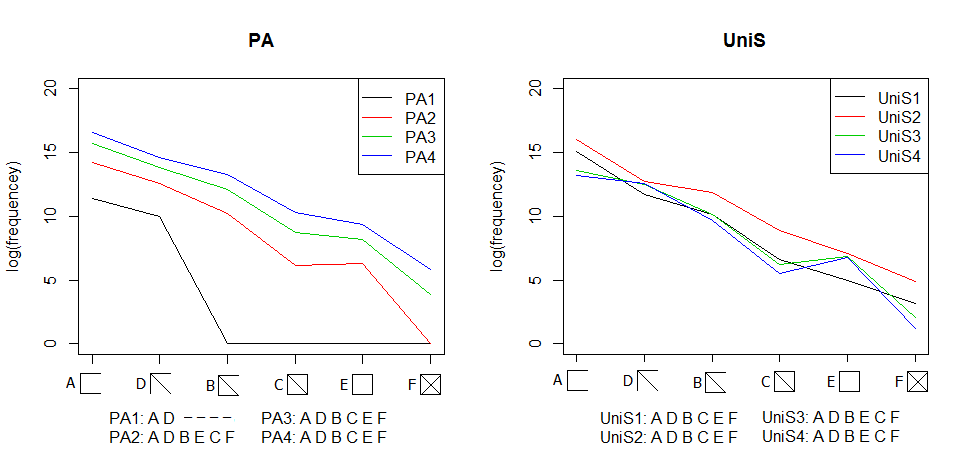}
\end{center}
\caption{Each figure shows the frequencies of six types of four-node motifs, for a  corresponding networks with $\textbf{m}\in\{1,2,3,4\}$. On the left, are the results for preferential attachment, on the right, the uniform sampling algorithm.  For each network, \PA{} or \UniS{} (denoted in the figure as UniS) with specified $m$, we compute the frequency of each of the size possible arrangements of four-node path-connected subgraphs (labelled A to F) and then plot the absolute (logarithmic) frequency of occurrence of each graph. The order from most to least frequent  is also provided.}
\label{motif}
\end{figure}

The motif, defined as a small connected subgraph that recurs in a graph, is the basic building block, or functional unit, of complex networks \cite{aJ09}. In real-world networks (e.g. gene regulatory networks), the relative frequencies of different motifs often represent different functions of the network \cite{rM02,uA07}. Moreover, it sheds light on the structure of the whole network. We restrict our analysis here to the four-node motifs, and classify them into three groups (based on the number of edges in the constituent motifs, see  Fig.~\ref{motif}): $\{A,D\}$, $\{B,E\}$, $\{C,F\}$. Measuring the relative frequency of each of these groups allows us to determine the proportion of building blocks typical of the sparse, moderate, and dense networks. The result of the motif analysis is also shown in Fig.~\ref{motif}.

From Fig.\ref{motif}, there is an apparent difference between the motif distribution of these two type of networks: as $m$ increases, the motif frequencies of PA networks become larger, and the ``dense'' motifs of type $C$ occur more frequently than the less dense type $E$ --- suggesting that the PA networks are denser than the corresponding \UniS{} networks. This would be a natural consequence of the PA networks becomes more hub-centric (more cross links between hubs) as $m$ increases, while the hubs of the \UniS{} network remain more evenly distributed (less cross-links between hubs, and correspondingly fewer ``dense'' motifs).

\subsection{Robustness and attack vulnerability}

Due to the long tail of the power-law distribution, scale-free networks
are often claimed to be simultaneously robust to the random loss of nodes
(i.e. ``error tolerance'') and fragile to targeted worst-case attacks
\cite{aR00}. The robustness is seen to be a
consequence of the extremely high number of (``unimportant'') low degree
nodes, the fragility is due to the extremely important role of the
hubs. { This property is also called the ``Achilles' heel'' of PA
networks, or the ``robust yet fragile'' feature}. Intuitively, the
inhomogeneous connectivity distribution of many networks caused by the
power-law distribution would possess these two properties. However, from
our analysis, \UniS{} networks generated by sampling uniformly from the
family of all scale free networks do not exhibit this second
property. Again, we may attribute this to the hub-centric nature of PA
networks.

\begin{figure}[!htb]
\begin{center}
\includegraphics[width=0.75\textwidth]{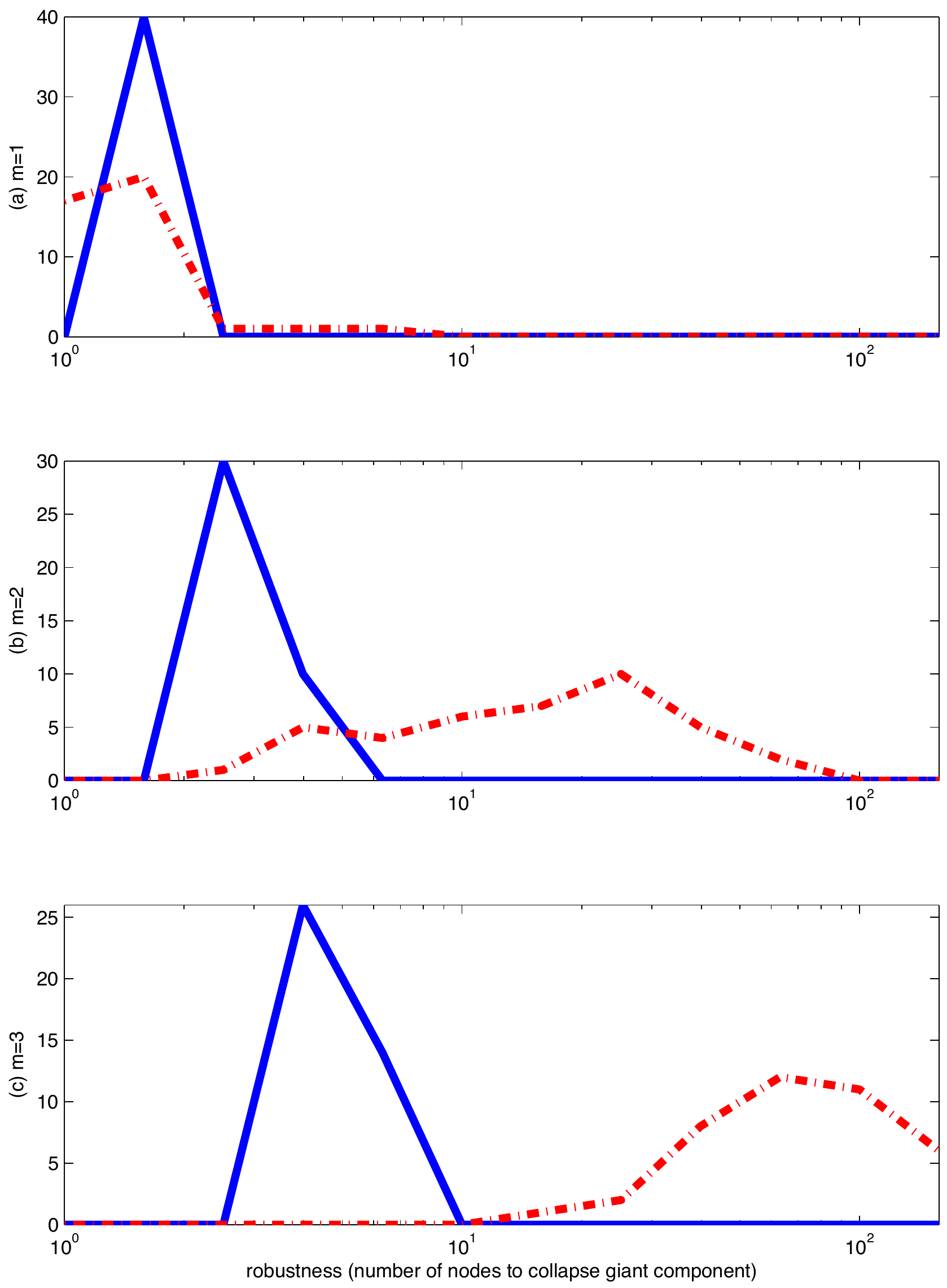}
\end{center}
\caption{Numerical estimates of histograms showing the number of nodes which must be removed to induce collapse of the network giant component. Nodes are selected for removal via either node degree or node betweenness centrality (here we illustrate results for removal via node degree --- results for betweenness centrality are visibly indistinguishable). Panels (from top to bottom) are for different minimum degrees $m=1,2,3$. Results for PA are plotted as a solid blue line and and results for \UniS{} are plotted as a red dashed line. Histograms are computed with logarithmically equally space bins --- the same binning is used for both network construction methods.
We find that, \UniS{} networks are, for $m>1$, an order of magnitude more robust to targeted attack than typical PA networks.}
\label{robu}
\end{figure}

We quantify the robustness to targeted attacks by specific removal of the most highly connected (or important) nodes until the network is disconnected. We then take the number of the nodes removed as a measure. For illustration, we use degree and betweenness centrality (BC) of a vertex as measures with which to target nodes for removal. Roughly, BC is defined as the number of geodesics (shortest paths)
going through a vertex: 
\[{\rm BC}(v)= \sum _{ i\neq j,i\neq v,j\neq v }^{  }{ { g }_{ ivj }/{ g }_{ ij } },\] where $g _{ ij } $ is total number of shortest paths between node $i$ and $j$, and $g _{ ivj } $ is the number of those paths that pass through $v$. The results are plotted in  Fig.~\ref{robu}.

The case of targeted attack is trivial for $m=1$, in which situation the networks are highly likely to be trees. Hence, we restrict our analysis for the case where $m=2,3$. these results are shown in Fig.~\ref{robu}.
From Fig.~\ref{robu}, it is safe to conclude that the \UniS{} network is much more robust than the corresponding PA network when facing targeted attack. This is a consequence of the fact that the PA networks are more hub-centric, while the \UniS{} networks are not. The fragility of the PA networks is due to the placement of hubs within the network --- not the scale-free-ness of the network {\em per se.}

\subsection{Numerical statistics of network}

\begin{figure}
\begin{center}
\includegraphics[width=0.95\textwidth]{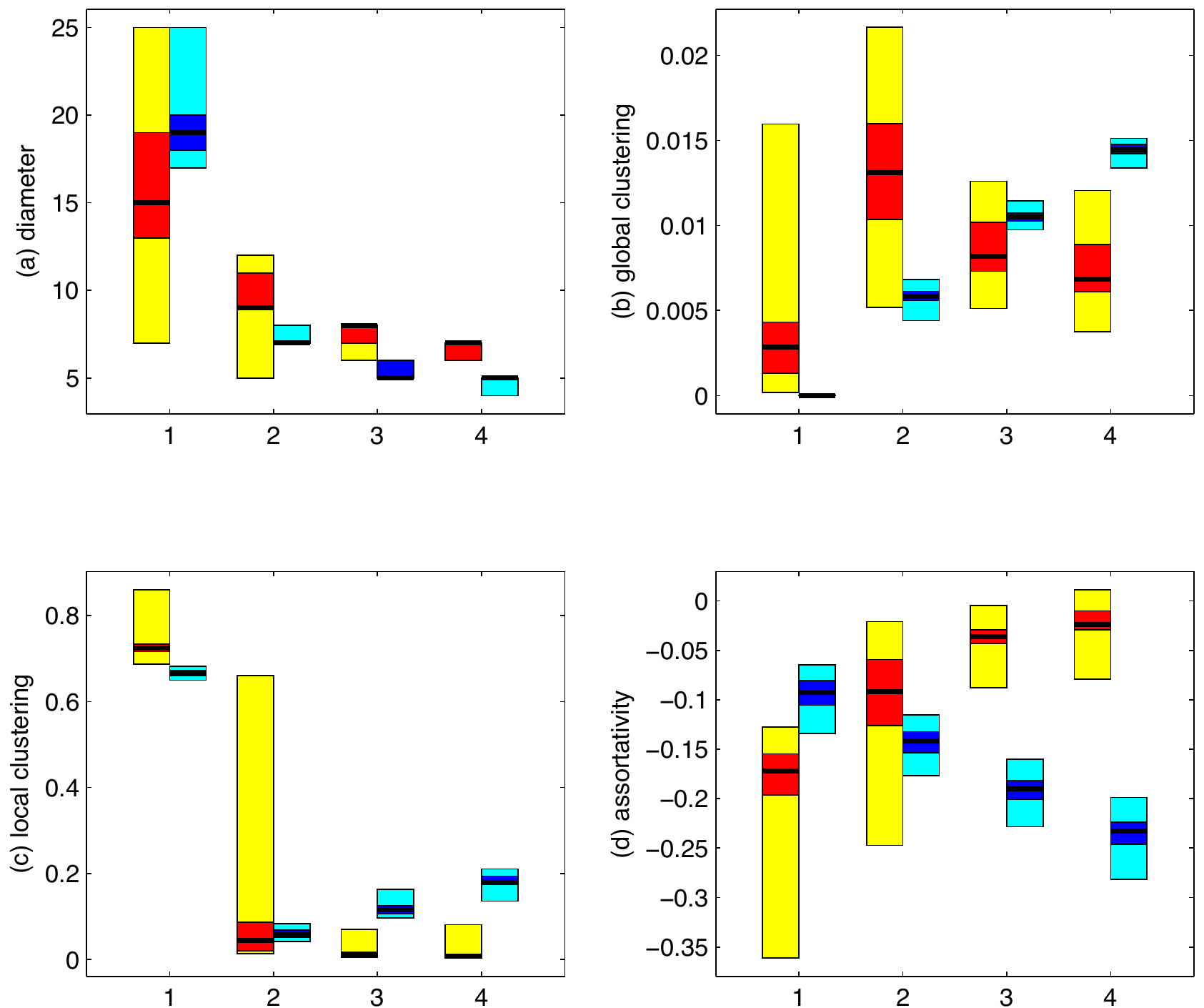}
\end{center}
\caption{Boxplot of the four network statistics described in the text { and Table \ref{tab:tab1}}. The boxplot depicts maximum and minimum ({extremum of the box), upper quantile ($75\%$) and lower quantile ($25\%$) (the darker box), and median (solid dark line within the inner box)} of the data. The { left-most box within} each pair is the summary of the \UniS{} network, while the right one represents PA networks.  Computation is depicted for $m=1,2,3,4$. Note that, for example, assortativity has an increasing negative bias for PA networks, while \UniS{} approaches $0$ as $m$ is increased.}
\label{box}
\end{figure}

In this section we present the difference between the PA networks and \UniS{} networks by computing the widely quoted numerical network properties:
\begin{itemize}
\setlength{\itemsep}{0pt}
\setlength{\parsep}{0pt}
\setlength{\parskip}{0pt}
\item \emph{diameter}: the maximal shortest path length.
\item \emph{global clustering coefficient}: the proportion of the number of closed triplets divided by the number of connected triples of vertices, which is a measure of degree to which nodes in a graph tend to cluster together.
\item \emph{local clustering coefficient}: the proportion of links between the vertices within its neighborhood divided by the number of links that could possibly exist between them, which quantifies how close its neighbors are to being a clique.
\item \emph{assortativity}: the Pearson correlation coefficient of degree between pairs of linked nodes, which measures the preference for a network's nodes to attach to others that are similar in degree.
\end{itemize}
The application of these four statistics is summarized in Fig.~\ref{box}.

\begin{sidewaystable}
  \centering
    \begin{tabular}{r|r|r|r|r|r}
    \toprule
          & diameter & global\_cc & local\_cc & assortativty & motif rank \\
    \midrule
    m=1 PA & 19.47 (1.71) & 0     & 0.67 (0.01) & -0.095 (0.017) & BECFDA \\
    m=1 \UniS{} & 15.55 (4.33) & 0.0037 (0.0033) & 0.73 (0.03) &  -0.18 (0.05) & FECBDA \\
    m=2 PA & 7.05 (0.22) & 0.0059 (0.0005) & 0.061 (0.011) & -0.14 (0.014) & FCEBDA \\
    m=2 \UniS{} & 9.45 (1.53) & 0.014 (0.004) & 0.093 (0.14) & -0.095 (0.052) & FECBDA \\
    m=3 PA & 5.28 (0.45) & 0.011 (0.0004) & 0.12 (0.014) & -0.19 (0.015) & FECBDA \\
    m=3 \UniS{} & 7.57 (0.54) & 0.0087 (0.0021) & 0.017 (0.014) & -0.038 (0.019) & FCEBDA \\
    m=4 PA & 4.93 (0.27) & 0.014 (0.0004) & 0.18 (0.017) & -0.24 (0.019) & FECBDA \\
    m=4 \UniS{} & 6.67 (0.47) & 0.0072 (0.0019) & 0.011 (0.011) & -0.021 (0.017) & FCEBDA \\
    \bottomrule
    \end{tabular}%
  \label{tab:tab1}%
    \caption{Summary of numerical statistics for the four pairs of networks ($m=1,2,3,4$). { Details of computation and comparison of statistical distributions are provided in the main text and relevant figures.}}
\end{sidewaystable}%

\begin{figure}
\begin{center}
\includegraphics[width=0.65\textwidth]{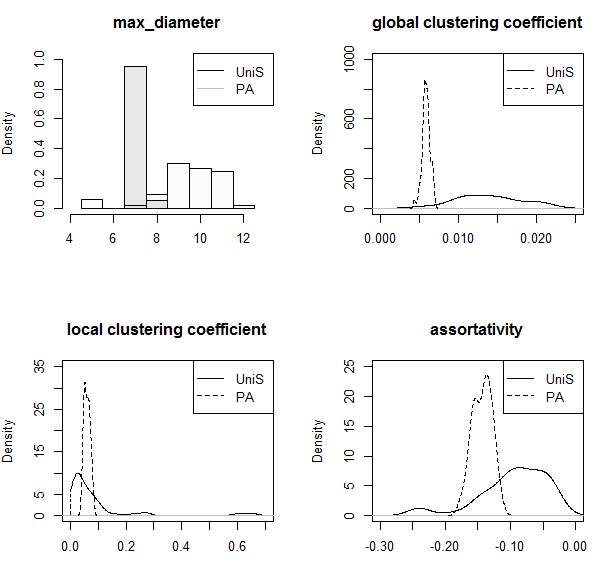}
\includegraphics[width=0.65\textwidth]{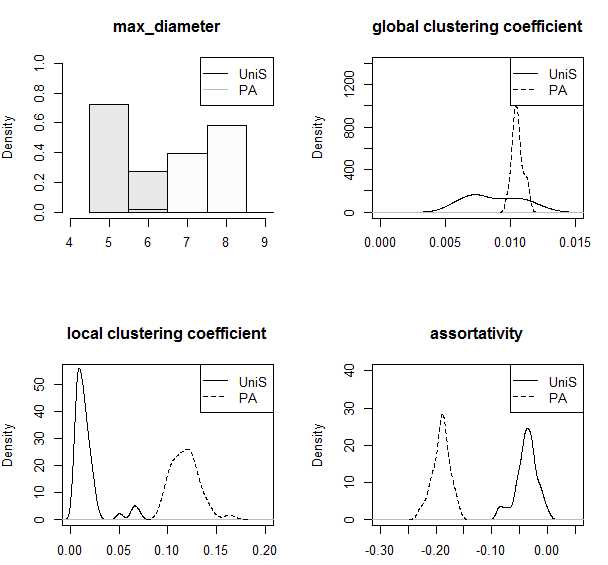}
\end{center}
\caption{Distribution of the statistics of networks with $m=2$ (upper four panels) and $m=3$ (lower four panels). Diameter is a straight histogram count, and the probability densities in the other panels are computed with a Gaussian kernel estimator. We see systematic bias and decreased variance for preferential attachment across all indicators --- exactly as one would expect. Most importantly, PA networks are smaller (diameter) typically more strongly dissassortative (see \cite{sJ10}). The \UniS{} networks are labelled in the figure captions as UniS --- for uniform random sampling.}
\label{stat2}
\end{figure}

We select cases with $m=2$ and $m=3$ to visualize the curve of these statistics for the pair of networks in Fig.\ref{stat2}. 
Although not shown, we note here that the case with $m=1$ is similar to the case $m=2$, while $m=4$ is  similar to $m=3$.

From the curves of  various network statistics depicted above, we find that PA networks are atypical. In particular, PA networks have: (i)  more negative assortativity compared with corresponding \UniS{} networks; (ii) increasing clustering coefficient as $m$ increases; and, (iii)  in the case $m=3$ and $m=4$ the two clustering coefficient curves separate from each other. All these  discrepancies point to the fact that as $m$ increases, the PA network becomes more and more hub-centric, while the \UniS{} network remain highly uniform -- and the high degree nodes  remain evenly distributed throughout the \UniS{} networks.

We now introduce algorithms to modify the specific aspects of the network
which contribute to this hub-centric property. Note that the hub-centric
structure is a global property of the network, since only modification on
a small portion of PA networks (such as the case with the so-called
``rich-club'' phenomena \cite{xu4}) is not sufficient: when we only
make the modification described in \cite{xu4} to manipulate rich-club
connections the PA and \UniS{} network statistical distributions remain
disparate. Therefore, the modification scheme we propose in the following
applies across the entire network structure --- from super-rich nodes and
hubs, to the poorest nodes. To maintain conciseness and focus, we present
the brief idea of our modifications scheme here; for details see
Algorithm~\ref{alg:spread} of  \ref{modification}. In the
following paragraphs we outline these two modification schemes --- one to
remove what we call the hub-centric features of PA networks and a second
scheme to add these features to \UniS{} realisations. The aim of these
computations is to show that these modifications alone are sufficient to
align the corresponding distributions of network structural properties
(assortativity and so on).

\begin{figure}
\begin{center}
\includegraphics[width=0.95\textwidth]{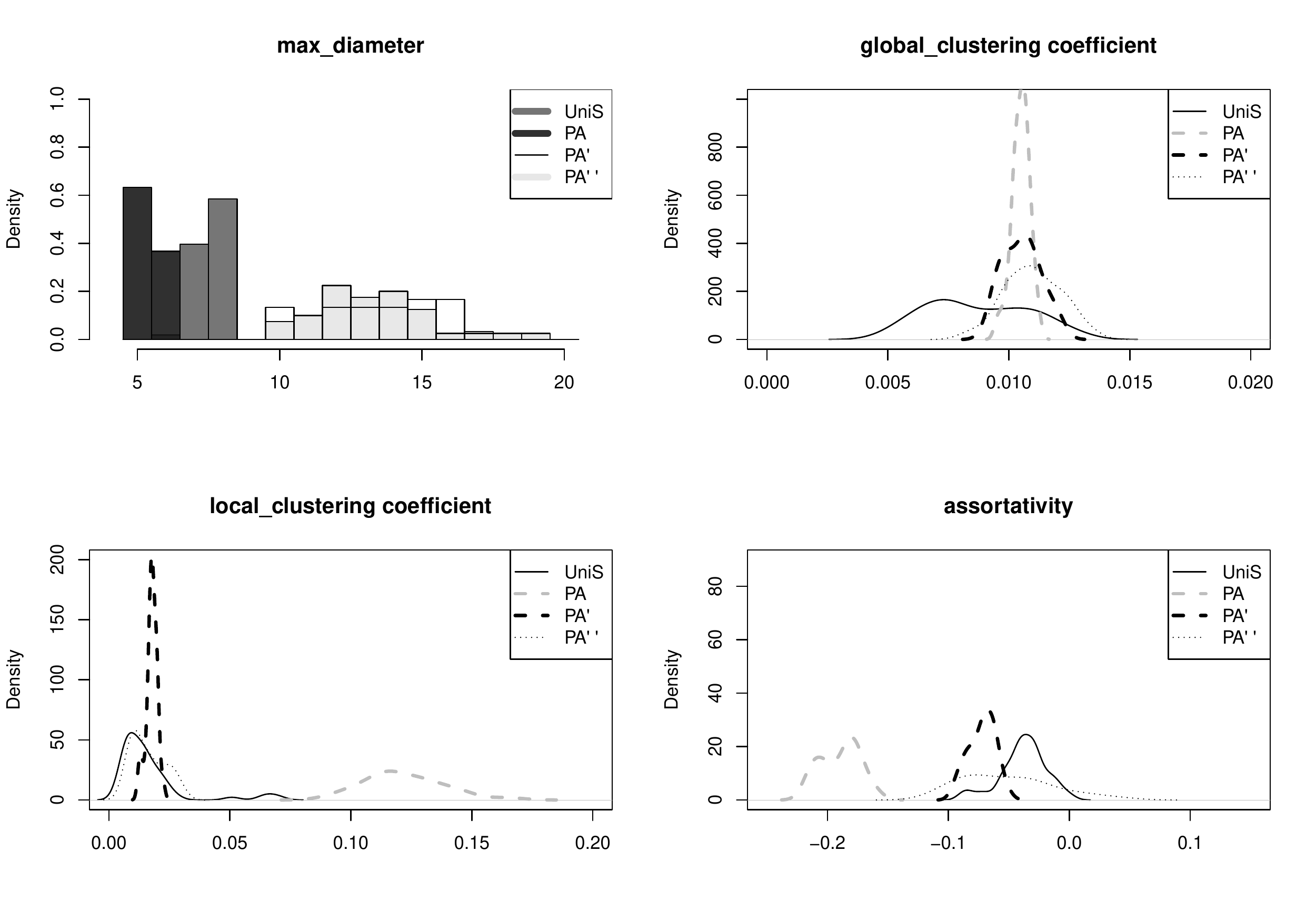}
\end{center}
\caption{(\textbf{Modification on PA}) The probability distribution function curve of the statistics for \UniS{} networks (labelled UniS, solid lines), PA networks (bold dotted gray lines), modified PA networks (termed PA\'\, with bold dotted black lines) and modified fitted PA networks (termed PA\'\,\'\,with dotted black lines), where $m=3$.}
\label{mod1}
\end{figure}

The modification scheme for an initial PA network is the following. First, we cut the links within the group of rich nodes and between rich nodes and the group of poor nodes as far as possible while preserving the connectivity of the network. Then, we use the algorithm described in \cite{xu4} among the giant nodes (which we define as the nodes whose degree is even larger than the typical rich nodes), to reconstruct the structure of the rich nodes. That is, we obtain networks with minimal interconnection between hubs, minimal connection from hubs to low degree nodes, but a similar rich-club structure (the interconnection among the super hubs).  After setting the thresholds $\alpha_1=60\%$, $\alpha_2=5\%$ and $\alpha_3=0.5\%$ in Alg.~\ref{alg:spread} (\ref{modification}), we apply this modification to the case with $m=3$ and $m=4$. The result with $m=3 $ is shown in Fig.~\ref{mod1}, and for conciseness, the result with $m=4$, although similar, are omitted.

From Fig.~\ref{mod1}, the range of the {{probability distribution functions}} for the modified graphs overlap with the unmodified target distribution \UniS{} --- indicating that our modification provides a good fit to the distribution of PA networks, with the {{addition}} of hub-centric links. Moreover, we test the modified curves of the PA networks and the curves of \UniS{} networks. For this purpose, we add the following random factors. First, we choose $\alpha_1$ randomly from the set $[50\%,80\%]$, and then the result for $m=3 $ is also depicted in Fig.~\ref{mod1}. The results for $m=4$ are similar, but omitted for conciseness.

From the calculations of Fig.~\ref{mod1} we see that the modification of the PA scheme (removing the hub-centric properties of such networks) allows us to produce network statistics which have a distribution similar to the unmodified uniform sampling scheme (for the local clustering coefficient and assortativity \cite{sJ10}, at least) or bracketing the expected distribution for \UniS{} (for the maximum diameter). Although this bracketing --- by modifying the hub-centric nature of the PA network we go from smaller than \UniS{} to larger that \UniS{} --- does not immediately provide indistinguishable statistical distributions, it is clear that this is due only to the unselective manner in which we choose the threshold parameters for this algorithm. Changing these parameters effects a continuous change in these statistical distributions. This is sufficient to make our case that these hub-centric groups of nodes are what causes the PA networks to be atypical.  We do note, however, that the parametric changes we have explored are insufficient to modify the PA network  to reproduce the full distribution of observed global clustering coefficients for the \UniS{} network: there appears to be still further unexplored richness in the variety of these networks --- scale-free networks with close to zero clustering (very tree-like networks \cite{kJ13}).

\begin{figure}
\begin{center}
\includegraphics[width=0.95\textwidth]{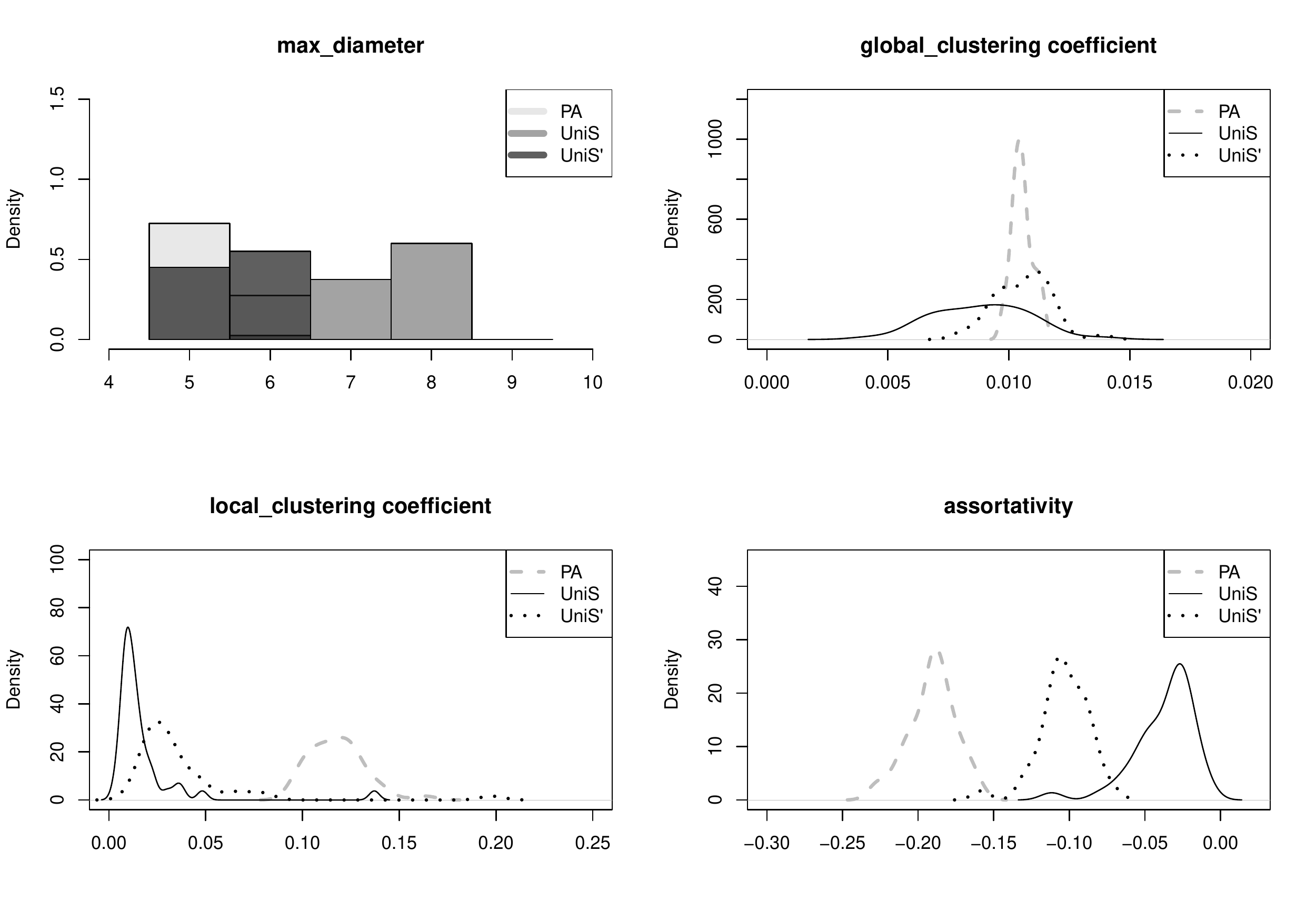}
\end{center}
\caption{(\textbf{Modification on \UniS{}}) The probability distribution function curve of the \UniS{} networks (labelled UniS with solid lines), PA networks (bold dotted gray lines) and modified \UniS{} networks (termed UniS' with dotted lines) in the case $m=3$}
\label{rev3}
\end{figure}

To further support our claim, we also present here a reverse modification
algorithm on \UniS{} networks to generate interconnection among hubs,
which would result in the adjusted curves of the statistics for \UniS{}
networks approaching that of the PA networks. Leaving the details to
\ref{modification} (Alg.~\ref{alg:concentrate}), we again
introduce the idea briefly here.  First, we delete the edges among the
poor nodes and relink these edges to rich nodes with probability
proportional to the degree of the rich nodes. Second, we use the
algorithm described in \cite{xu4} to make a club of super-rich
(``giant'') nodes (the definition is similar to the one in the first
modification) connected. In this way we are able generate the hub-centric
structure in networks. Setting $\alpha_1=80\%$, $\alpha_2=5\%$ and
$\alpha_3=0.5\%$ for Alg.~\ref{alg:spread}, the result for $m=3$ is shown
in Fig.~\ref{rev3}. The calculation for $m=4$ is similar and omitted for
conciseness.

From Fig.~\ref{rev3}, after generating hubs by artificially reconstructing the structure of networks, the numerical analysis of \UniS{} networks indicates statistical distributions of structural properties which approach that of PA networks --- supporting our claim that hub-centric structure in PA networks makes them atypical of random scale-free graphs.

By modifying the properties of a uniformly sampled network we are able to
produce distributions of maximum diameter and global clustering
coefficient very similar to that observed for the PA. That is, we add
hub-centric structure to  \UniS{} network and achieve statistical
features similar to that typical of PA networks. We do not achieve such
good agreement for assortativity and local clustering
coefficient. Nonetheless, taken together Fig.~\ref{mod1} and~\ref{rev3}
show that matching distributions can be achieved by {\em either} adding or
removing the hub-centric features in each of the four statistics that we
examine here. Hence, the features which we modify with the algorithms
described in \ref{modification} are exactly the properties of PA networks that
are atypical of the broader distribution of properties of scale free networks
when sampled by a maximum entropy process. The only caveat being that the
two algorithms are not exactly inverses of one another, and, in
particular, measures of global clustering hint at further unexplored
diversity in the global structure of typical scale free networks (as the
networks become progressively more tree like, even for $m>1$).

\subsection{Hub-centric structure in PA networks}

To conclude this section, we now collate the results of our analysis.  We
see that the hubs of a PA network ``hold'' that network together.  The
strength of interconnection among hubs (and connection with low degree
nodes) may be explained by preferential attachment itself. In that
growth model, once nodes and links are added, they will never be altered
again. This, we claim, causes inhomogeneity in PA networks. The largest
hubs will always (with probability approaching one) be the earliest nodes
in the growing network and these nodes are necessarily closely
interlinked. Conversely, the last nodes added to the network will have
minimal degree and yet these nodes will (with very high probability) be
directly connect to the hubs. In fact, it is clear that the last nodes
added (those with the lowest degree) are {\em most} likely to be connected
directly to the hubs  --- this is a consequence of the fact that the
  attachment bias favouring the original rich nodes increases with growth
  of the network --- loosely speaking, that the ``rich get richer''. These
are precisely the properties we alluded to earlier with our description of
``hub-centric'' networks, and, these are precisely the properties which
are adjusted by the modification algorithms~\ref{alg:spread} and~\ref{alg:concentrate}
of \ref{modification}.

The consequences of this hub-centric structure of PA networks is two-fold. {\em First}, since PA is an elegant and intuitive way to generate graphs, there may exist some generation procedures in the real world which are similar to preferential attachment, and thus we can use this claim to illustrate the hub-centric structure in such growth networks. This observation has potential for practical use --- for example, with hub detection in control of disease transmission, or,  to control of a network by manipulating the hubs.
{\em Second}, such a claim also indicates that there is systematic bias with PA networks. This bias will lead to difficulty when using PA networks to explain real world networks which do not result from such a constrained growth process --- even when the degree sequence of each network satisfies the power law distribution.

\section{Surrogate Networks: An application to particular putative scale-free systems}
\label{sec:applications}

In this section, we introduce a variant of the surrogate data test,
proposed for nonlinear time series analysis, to interpret real world
networks. We proceed by generating an ensemble of random networks, both
similar to the observed data-based network (in that they share the same
degree distribution) and yet, at the same time, random. By comparing the
properties of the real network with the corresponding distribution of
\UniS{} networks we can determine whether the particular network is
typical. Our observations here are both a consequence of the previous
section and a motivation for the development of a network surrogate test. 

Just like the surrogate data methods for time series, results will depend on the choice of test statistic. We will see that certain test statistics indicate clear and systematic deviation between the maximum entropy realisations and particular experimental systems. This can be interpreted in two ways. {\em First}, as a straight-forward hypothesis test. Rejection on the basis of any particular statistic indicates that the observed network is not typical of the maximum entropy class. However, we are also able to be more particular. Hence, {\em second}, we can apply this method to determine which particular features of the observed network are atypical. Hence we provide a more sensitive classification scheme dependent on individual features of the network. In what follows we will see that this is very often the case --- it is particular features of the networks that differ, rather than {\em all} aspects {\em simultaneously.}

We will start with a brief discussion of robustness. As we saw above, PA
networks are vulnerable to targeted attack, while \UniS{} networks don't
have such an evident ``Achilles' heel''. Recent research into the
structure of several important complex networks shows that, even if their
degree distribution could have an approximately power-law tail, the
networks in question are robust to targeted attack to some degree: the most
highly connected nodes do not necessarily represent an ``Achilles' heel''. In particular, recent results of modeling the router-level
Internet has shown that the core of that network is constructed from a
mesh of high bandwidth, low-connectivity routers, and \cite{jD05b}
concludes that although power-law degree distributions imply the presence
of high-degree vertices, they do not imply that such nodes form a
necessarily ``crucial hub''. A related line of research into the structure
of biological metabolic networks have shown that claims of SF structure
fail to capture the most essential biochemical as well as ``robust yet
fragile'' features of cellular metabolism and in many cases completely
misinterpret the relevant biology --- for example, \cite{rT05}
indicates the removal of high degree nodes leaves the biosynthetic
pathways fully intact. Hence, real-world scale-free networks do, indeed,
exhibit absence of the much-touted ``robustness'' properties. Our model
provides an explanation for that absence.

In the following, we will probe the limitation of the explanative power of PA networks via numerical statistics. We find that  many real complex networks appear more ``uniform'' under our surrogate test. Also, based on this result, we propose a simplistic classification of real world networks. Such a classification could provide us with a new insight into the interior structure of real networks, to explore which property do make the networks interesting and special.

\begin{figure}[!htp]
\begin{center}

\includegraphics[width=0.95\textwidth]{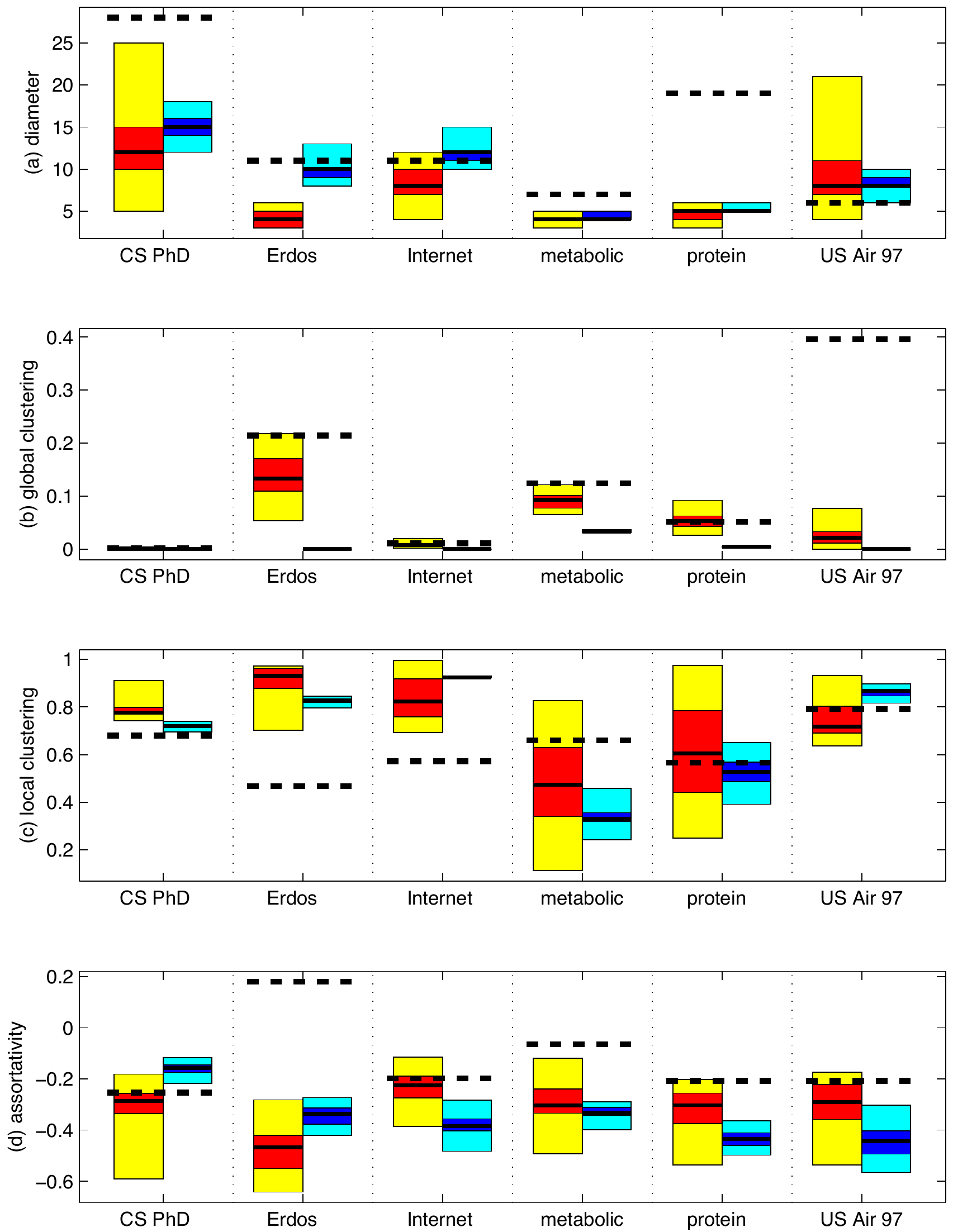}
\end{center}
\caption{Network characteristics: (a) diameter; (b) global clustering coefficient; (c) local clustering coefficient; and, (d) assortativity. Boxplot analysis for collaboration and information networks, depicting  maximum, minimum, upper quantile ($75\%$), median and lower quantile ($25\%$) of the data.  $1.$ CS PhD collaboration\, $2.$ Erd\"os collaboration \, $3.$ a symmetrized snapshot of the structure of the Internet at the level of autonomous systems\, $4.$ a metabolic pathway network,  $5.$ the {\em S.cerevisiae} protein-protein interaction network\, $6.$ US Airport connection.}
\label{box1}
\end{figure}

Our surrogate tests are developed in this way: for a real world network
which satisfies the power-law distribution, we estimate its minimum vertex
degree and power law exponent by using the algorithm described in
\cite{aC09}. We then use the relation $\gamma=2+A/m$ in {Section
  $1$} to estimate its initial attractiveness, and then generate PA
networks with controlled $\gamma$ and $m$. We then substitute $\gamma$ and
$m$ into Algorithm~\ref{alg:MEp} to generate the \UniS{} networks. For
each set of parameter values, we generate $40$ networks for \UniS{} and PA
networks respectively.  In Fig.~\ref{box1}, we draw the boxplot of
numerical statistics of the real networks, as well as its corresponding PA
and \UniS{} networks.

\begin{figure}[!hp]
\begin{center}
\includegraphics[width=0.65\textwidth]{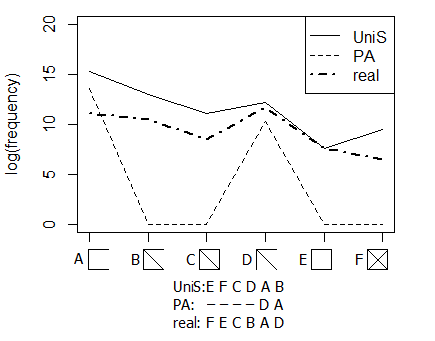}
\includegraphics[width=0.90\textwidth]{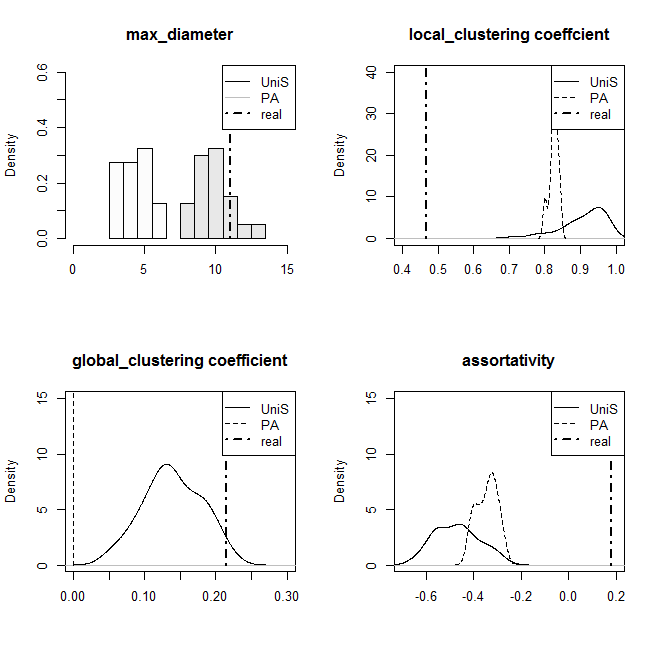}
\end{center}
\caption{The first panel shows the motif distribution and the lower panels the curves of statistics of the \UniS{} networks (solid lines, labelled UniS), PA networks (dotted lines) and  Erd\"os collaboration network (bold mixed lines). The motif rank in ascending order is shown below the graph.}
\label{erdosmotif}
\end{figure}

From Fig. \ref{box1} we can conclude that, in terms of interpretation of real networks, \UniS{} networks provides a significantly better model for  data from the Internet topology, and metabolic and protein interaction networks. For CS PhD collaboration and US airport connection networks, \UniS{} is slightly better than PA. However, for Erd\"os collaboration network, neither \UniS{} nor PA networks is good, although the statistics of PA networks are closer to that of the Erd\"os network than \UniS{} networks.

To examine the case of Erd\"os collaboration network more closely, we plot the motif rank and various other statistics of the Erd\"os collaboration networks in Fig. \ref{erdosmotif}. 
This figure suggests that the hub-centric structure in the Erd\"os collaboration network is much more significant than PA allows: the Erd\"os collaboration network is a densely connected core along with loosely coupled radial branches reaching out from that core.  Erd\"os practiced what he preached -- he was a weaver of social networks and thus a builder of social capital. Moreover, this collaboration network is specifically Erd\"os-centric --- it is specifically focussed on that one unique hub and its connections. Hence, the nodes closer to Erd\"os benefit most and become the strongest hubs (after Erd\"os himself) in the resultant network.

By comparison, it would be helpful for us to identify which properties differ most markedly between the real network and a random surrogate sample.. The cases where \UniS{} provides better representations would suggest, for
instance, that the structure of protein interaction networks is more
uniform, because the cellular ecosystem necessitates such stableness; and,
in the case of the the Internet on the level of automatic system, the
system is balanced and distributed in a rather deliberate way. Both these
systems have been ``engineered'' for robustness.  Finally, we note that
neither the \UniS{} nor PA networks performs well in reproducing the
diameter of the biological networks. It is curious that biological
networks present abnormally large diameters --- we speculate that this
more spread-out tree-like structure is a consequence of the components of
biological systems having specific functional roles.

Based on our results, we also note (via analysis of robustness, motif and numerical statistics of a network, and comparison with its corresponding PA and \UniS{} networks) that one can classify the real network in the sense of hub-structure: if its property is closer to the PA networks, we can conclude it is hub-centric and typically the result of a growth process. Conversely, networks that are not PA are more likely to be purposefully widely distributed.

\section{Conclusion and discussion}
\label{conclude}

We have proposed a new algorithm which allows us to make maximum-entropy
random-samples of finite size graphs with the degree histogram being a
probabilistic realization of a specified degree distribution.  We focus
on  putative scale free networks as an illustrative example and use this method to generate random networks
with power-law tail degree distributions of arbitrary $\gamma$ (exponent)
and $m$ (minimum degree). This provides a simple alternative to the
various generative processes in the literature. However, our approach has
the benefit that it makes no particular biasing assumptions (such a
preferential attachment). To emphasize the need for this algorithm we
compare our results to distributions of networks obtained from the widely
applied PA scheme of \cite{aB99}.  While there are many generative algorithms for scale-free networks, PA is still the most widely used. Hence, PA provides an excellent model of
growth of a static network, however, we find that many real world networks
do not conform to this model.  This is not a new observation, however, it is still an important point to emphasise --- growth (and preferential attachment in particular) are not adequate models to explain all observed scale-free networks. We  propose particular structural modification algorithms to uncover exactly how PA is atypical of what one would expect of maximum entropy.

We found the high-degree nodes in PA networks are hub-centric, and that this hub-centric-ness has a greater influence on the overall structure than attributable only to the high-degree nodes themselves. In particular we attribute the co-called ``robust yet fragile'' feature found in many scale free networks to the distribution of hubs achieved via preferential attachment. Our observation sod the hub-centric-ness of preferential attachment scale free networks helps us assess the contribution of hubs in real world networks to the overall network structure. The surrogate network generation method we describe allows us to provide a robust numerical measure of this rich variation among scale-free complex networks.

\appendix

\section{Implementing ${ME(\p)}$}\label{App:mainalg}

A \emph{maximum entropy} process~${ME(\p)}$ can be implemented by as a
Monte Carlo Markov Chain (MCMC)~\cite{kJ13}. However, a naive MCMC
implementation could converge slowly and not be effective for large
graphs~\cite{pD12, pD13, yC05}. Here we state an alternative
algorithm; the key ingredients being \emph{edge
  switching}~\cite{aS11, rT81}, connectivity testing~\cite{kJ13},
and constructing graphs by the Havel-Hakimi process~\cite{vH55, sH62}. {{}}

For $G\in\G$ let $\d(G)=(d_1,\dots,d_N)$ denote the \emph{degree
  sequence}, that is, $d_i>0$ is the degree of node~$i$. An arbitrary
degree sequence $\d\in\Z^{N}$ is called \emph{graphical} if there exists
$G\in\G$ such that $\d(G)=\d$.

\begin{algorithm}\label{alg:MEp}
  Let $\p$ be a target probability mass (with power-law tail).
  \begin{enumerate}
  \item Generate a sample $\n$ from the multinomial distribution of~$\p$.
  \item Generate a uniformly sampled random degree sequence~$\d$
    from~$\n$.
  \item Use the Havel-Hakimi process to test $\d$ is graphical and
    construct the canonical graph~$G$ with $\d(G)=\d$. If $\d$ is not
    graphical, then return to step~2.
  \item Test whether $G$ is connected; if not, then return to step~2.
  \item Apply edge-switching to~$G$ to uniformly sample the equivalence
    class of $\n(G)$.
  \end{enumerate}
\end{algorithm}

Although this algorithm is generally efficient for large graphs, when
$m=1$ and $\gamma$ sufficiently large ($\gamma>\gamma_1$, \cite{kJ13})
the Havel-Hakimi process often constructs disconnected graphs, and so
step~3 becomes a bottle-neck. Fortunately, in this case an MCMC
algorithm~\cite{kJ13} which employs edge switching~\cite{aS11,
  rT81} is efficient.

Step~1 requires a sample~$\n$ from the multinomial distribution of~$\p$,
and step~2 a uniformly sampled degree sequence~$\d$ for~$\n$. Let
$q_k=\sum_{i=1}^kp_i$, $q_0=0$, and $I_k=[q_{k-1},q_k]$. Choose uniform
random variants~$x_i\in(0,1)$, $i=1,\dots,N$. Then let $d_i=k$ if
$x_i\in{}I_k$ and $n_k=|\{x_i\in{}I_k\}|$.  With step 3 we must determine whether a given $\d$ is
graphical. Havel~\cite{vH55} and Hakimi~\cite{sH62} independently
developed a test of $\d$ being graphical. The following Havel-Hakimi test
implicitly constructs a canonical graph~$G$ if $\d$ is graphical. Let
$N\geq2$ and $\widehat\d=\d$.
\begin{enumerate}
\item Choose $i$ such that $\widehat{d}_{i}>0$.
\item If $\widehat\d$ does not have at least $\widehat{d}_{i}$ entries
  $\widehat{d}_{j}>0$, $j\not=i$, then $\d$ is not graphical.
\item Subtract $1$ from the $\widehat{d}_{i}$ entries $\widehat{d}_{j}$,
  $j\not=i$, of highest degree. Set $\widehat{d}_{i}=0$.
\item If $\d=0$, then $\d$ is graphical otherwise return to step~1.
\end{enumerate}
We note here that there is another efficient method for graphicality~\cite{cG10}, but the Havel-Hakimi is sufficient for our implementation.
The canonical realization of $G\in\G$ for a graphical~$\d$ is implied by
step~3 of the test: the node~$i$ is connected to those other nodes~$j$
selected in step~3. The graph~$G$ can be built as the test proceeds by
constructing its adjacency matrix~$A$.

Step~4 of Algorithm~\ref{alg:MEp} requires one to test whether a graph~$G$ is
connected. This could be done by applying Dijkstra's algorithm on a certain node. 
Step~5 requires modifying a graph~$G$ by \emph{edge switching}. Let $A$ be
the adjacency matrix of~$G$. Let $i$, $j$, $k$, $l$ be distinct nodes,
such that $A_{ij}=A_{kl}=1$ and $A_{il}=A_{kj}=0$. Then the edges are
switched by setting $A_{ij}=A_{kl}=0$ and $A_{il}=A_{kj}=1$. Edge
switching does not change $\n(G)$, and if repeated sufficiently often the
resulting graph is  uniformly sampled from its equivalence
class~\cite{rT81,sK75}.

\section{Modification algorithms}
\label{modification}

In this section we provide the detailed modification algorithms described in
the main text. These algorithms are presented here to separate them from
the more central \UniS{} network generation algorithm (Alg.~\ref{alg:MEp}).  The first
modification algorithm (Alg.~\ref{alg:spread}) changes the distribution of PA
networks' statistics, by ``spreading hubs'', to make it closer to the \UniS{}
networks. Moreover, after adding random factors on the first modification,
we can fit the curve of the \UniS{} networks. We also present the contrary
modification algorithm (Alg.~\ref{alg:concentrate}) to change the curve of \UniS{}
networks to approach that of PA networks by ``concentrating hubs''.
 
 The following algorithm decentralizes hubs by spreading the hubs of a
 network.

\begin{algorithm}\label{alg:spread}
  \begin{enumerate}
  \item Start with a simply connected network $G$ (presumably generated
    with a PA process).
  \item\label{step1} Select three percentages $\alpha_1, \alpha_2,
    \alpha_3$ for the definition of the \emph{poor}, \emph{rich} and
    \emph{giant} nodes in the algorithm.
  \item\label{step2} Sort the degree sequence, select the $\alpha_1$
    lowest degree, and define the corresponding vertices as the {poor}
    nodes. Similarly, select the $\alpha_2$ and $\alpha_3$ highest degree,
    define them as the {rich} and {giant} nodes.
  \item\label{step3} Delete all the links between the {poor} nodes and
    {rich} nodes.
  \item\label{step4} Add the minimal number of links among the {rich}
    nodes to make them connected, and define them as a ``{club}''.
  \item\label{step5} Check the {poor} nodes sequentially, if one is
    not linked to the {club}, random choose one member in {club} and
    link this member to the {poor} node and add the poor node to the
    {club}, until the members of {club} includes all the vertices of $G$.
  \item\label{step6} Randomly pick up $2$ linked {giant} nodes $g_1,g_2$,
    and $2$ linked {non-giant} nodes $v_1,v_2$ which are not linked to 
    $g_1,g_2$.
  \item\label{step7} Apply the edge-switching method among
    $v_1,v_2,g_1,g_2$ \cite{xu4}. (At least one of the possible edge
    switches will result in a connected graph.)
  \item\label{step8} Repeat Step~\ref{step6} and Step~\ref{step7} until
    there is no links among the {giant} nodes.
  \end{enumerate}
\end{algorithm}

In brief, Steps \ref{step3}-\ref{step5} cut the links between the group of rich nodes and group of poor nodes, and Step \ref{step6}-\ref{step8} change the structure among the rich nodes, since it is this group of rich nodes possess most of the edges. Steps \ref{step6}-\ref{step8}  preserve the degree sequence. The result is that  the PA networks become less hub-centric. We also note that, in Step \ref{step4}, there is often no need to add links, since the group of rich nodes is often already connected.

We choose $\alpha_1=60\%$, $\alpha_2=5\%$, $\alpha_3=0.5\%$ for our modification algorithm, and apply it to the case with $m=3$ and $m=4$. The result with $m=3 $ is shown in Fig.~\ref{mod1}, other results are similar, but omitted for conciseness.

The following algorithm concentrates hubs by modifying the initial network
to make the hubs more strongly centralized.

\begin{algorithm}\label{alg:concentrate}
  \begin{enumerate}
  \item Start with a simply connected network $G$ (presumably generated
    with the \UniS{} process).
  \item\label{btep1}Select three percentage $\alpha_1, \alpha_2, \alpha_3$
    for the definition of the \emph{poor} \emph{rich} and \emph{giant}
    nodes in the algorithm.
  \item\label{btep2}Sort the degree sequence, select the $\alpha_1$ lowest
    degree, and define the corresponding vertices as the {poor}
    nodes. Similarly, select the $\alpha_2$ and $\alpha_3$ highest degree,
    define them as the {rich} and {giant} nodes.
  \item\label{btep3} Traverse the poor nodes of $G$.
    \begin{enumerate}
    \item For each poor node $v_i$, if there exists another poor node
      among its adjacent nodes, then select randomly one node among them,
      and delete the links between them.  Otherwise go to Step
      \ref{btep4}.
    \item For $v_i$, select one node from the group of rich nodes with the
      assigned probability proportional to their degree, link $v_i$ with
      it.
    \end{enumerate}
  \item\label{btep4}Randomly pick up $2$ not linked to {giant} nodes
    $g_1,g_2$, and $2$ not linked to {non-giant} nodes $v_1,v_2$ which are
    connected with $g_1$ and $g_2$.
  \item\label{btep5} Apply the edge-switching method among
    $v_1,v_2,g_1,g_2$ \cite{xu4}. (At least one of the possible edge
    switches will result in a connected graph.)
  \item\label{btep6} Repeat Step \ref{btep4} and Step \ref{btep5} until
    the subgraph induced by {giant} nodes is complete.
  \end{enumerate}
\end{algorithm}
This reverse modification algorithm on \UniS{} networks aims to generate interconnected hubs. Step \ref{btep3} deletes the edges between the  poor nodes and rich nodes as far as possible under the constraint of connectivity, and relinks these edges among rich nodes, and Step \ref{btep4} forces the giant nodes to connect with each other. Hence, we can generate the hub-centric structure in networks, which results in the adjusted curves of the statistics for \UniS{} networks approaching that of the PA networks.

Setting $\alpha_1=80\%$, $\alpha_2=5\%$ and $\alpha_3=0.5\%$ on this modification, the result for $m=3$ is shown in Fig.\ref{rev3}.

\section*{Acknowledgements}

MS is supported by an Australian Research Council Future Fellowship (FT110100896) and Discovery Project (DP140100203). LZ was supported by the UWA-USTC Research Training Programme.


\end{document}